\documentclass[preprint,showpacs,preprintnumbers,amsmath,amssymb]{revtex4}

\usepackage{graphicx}    % Include figure files
\usepackage{dcolumn}     % Align table columns on decimal point
\usepackage{bm}          % bold math

\begin{document}

\newcommand{\Ept}{\epsilon_{p_t}}

\newcommand{\Epp}{\epsilon_{p_2}}

\newcommand{\Ep}{\epsilon_p}

\newcommand{\be}{\begin{equation}}

\newcommand{\ee}{\end{equation}}

\newcommand{\scaleE}{E_S}

\title{Fusion reactions in plasmas as probe of the high-momentum tail of particle distributions}

\author{Massimo Coraddu}
\email{massimo.coraddu@ca.infn.it}
 \affiliation{Ist. Naz. Fisica Nucleare (I.N.F.N.) Cagliari,
             I-09042 Monserrato, Italy}
\affiliation{Dipart. di Fisica dell'Universit\`a di Cagliari,
             I-09042 Monserrato, Italy}
\author{Marcello Lissia}
\email{marcello.lissia@ca.infn.it}
  \affiliation{Ist. Naz. Fisica Nucleare (I.N.F.N.) Cagliari,
            I-09042 Monserrato, Italy}
\affiliation{Dipart. di Fisica dell'Universit\`a di Cagliari,
             I-09042 Monserrato, Italy}
\author{Giuseppe Mezzorani}
\email{giuseppe.mezzorani@ca.infn.it}
 \affiliation{Dipart. di Fisica dell'Universit\`a di Cagliari,
             I-09042 Monserrato, Italy}
 \affiliation{Ist. Naz. Fisica Nucleare (I.N.F.N.) Cagliari,
            I-09042 Monserrato, Italy}

\author{Piero Quarati}
\email{piero.quarati@polito.it}
 \affiliation{Dipartimento di Fisica, Politecnico di Torino,
   C.so Duca degli Abruzzi 24, I-10129 Torino, Italy}
\affiliation{Ist. Naz. Fisica Nucleare (I.N.F.N.) Cagliari,
            I-09042 Monserrato, Italy}
\date{November 30, 2005}      %  but any date may be explicitly specified

\begin{abstract}

In fusion reactions, the Coulomb barrier selects particles from 
the high-momentum part of the distribution. Therefore, small 
variations of the high-momentum tail of the velocity distribution 
can produce strong effects on fusion rates. In plasmas several 
potential mechanisms exist that can produce deviations from the 
standard Maxwell-Boltzmann distribution. Quantum broadening of 
the energy-momentum dispersion relation of the plasma 
quasi-particles modifies the high-momentum tail and could explain 
the fusion-rate enhancement observed in low-energy nuclear 
reaction experiments.

\end{abstract}

\pacs{}                      % PACS, the Physics and Astronomy
\maketitle

\section{\label{sec:intro} Introduction }
Many-body collisions broaden the relationship between energy and
momentum of quasi-particles: a distribution of momenta, which can
have long tails, characterizes a quasi-particle with a given 
energy. Therefore, the momentum distribution can be very 
different from the one obtained using a sharp correspondence 
between energy and momentum~\cite{Ga:67}. Plasmas are typical 
environments where this effect can be important.

Fusion processes select high-momentum particles that are able to 
penetrate the Coulomb barrier and are, therefore, extremely 
sensitive probes of the distribution 
tail~\cite{Coraddu:1998yb,St:00,St:02,Lissia:2005em}.

This broadening of the interacting particle energy-momentum
dispersion relation has been proposed recently
\cite{Starostin:2003next, Starostin:2003next2} as a possible
explanation  of the strong enhancement of the observed low-energy 
rate of the reaction $d(d,p)t$ in deuterated metal 
target~\cite{Raiola:02,Raiola:02a,Bonomo:03,Raiola:04,Raiola:05}.

In this paper we study the details of this quantum broadening
effect using a simple and effective expression for the 
distributions. In particular, we determine the region of the 
distribution responsible of the effect. Our method is applied to 
the specific case of the  enhancement observed in the $d(d,p)t$ 
reaction rate.

\section{Charged particle distribution in plasma}

We consider two species (1 and 2) of interacting charged particles
with mass, velocity, momentum, energy and density: $m_{1,2}$ ,
$\mathbf v_{1,2}$ , $\mathbf p_{1,2}$ , $E_{1,2}$ , $n_{1,2}$ .
Their fusion reaction rate is $r=(1+\delta_{12})^{-1} n_1 n_2
\langle \sigma v_{rel}\rangle$ where the reaction rate per
particle is:
\begin{equation}
   \label{eq:ChargPartReactRate}
    \langle  \sigma v_{rel} \rangle
     =    \int d^3\mathbf p_1  \int d^3\mathbf p_1\,
                                                  \Phi_1(\mathbf p_1)\, \Phi_2(\mathbf p_2)
                                                  \, \sigma v_{rel} \quad
                                                  ;
\end{equation}
$\Phi_{1,2}(\mathbf p)$ are the momentum distributions of
particles 1 and 2 and $ v_{rel} = |\mathbf v_1 - \mathbf v_2|$ is
their relative velocity.

The charged-particle fusion cross section $\sigma$ is conveniently
expressed as
\begin{equation}
    \sigma(\epsilon_p) = \frac{S(\epsilon_p)}{E}
                         \exp \left( - \sqrt{\frac{E_G}{\epsilon_p}} \right) \quad ,
    \label{eq:ChargPartCrossSec}
\end{equation}
where $S(\epsilon_p)$ is the astrophysical factor as function of 
$\epsilon_p = \frac{1}{2} \mu v_{rel}^2 = \frac{p_{rel}^2}{2\mu}$ 
with $\mu$ the reduced mass and $ E_G = 2 \mu c^2 (Z_1 Z_2^2 
\alpha \pi)^2 $ the Gamow energy. Note that the cross section 
depends on the relative momentum $p$ (to stress this point we 
write $\epsilon_p$); only in special cases there is sharp 
relation between the energy of the particle $E$ and $\epsilon_p$, 
for instance for free particles $E=\epsilon_p$ is the kinetic 
energy in the center of mass.

In general we may assume a relation of the form \cite{Ga:67}:
\begin{equation}
\label{eq:GenDispRel}
    \delta_\gamma (E, \Ep) = \frac{1}{\pi} \frac{\gamma}{(E - \Ep)^2 + \gamma^2}\quad ,
\end{equation}
where the width $\gamma = \hbar \nu_{coll}$ depends on the
collision frequency $\nu_{coll} = n \sigma_{coll} v_{coll} $.

Even when the energy distribution is Maxwellian $ \propto \exp
(E/k_bT)$ (we set the Boltzmann constant $k_b=1$), the resulting 
momentum distribution
\begin{equation}\label{eq:MomDistrDef}
       \Phi(\epsilon_p) d\epsilon_p =
        \frac{4\pi p^2 dp \int^{\infty}_{0} dE\, \delta_\gamma (E, \Ep) e^{-E/T} }
                       {4\pi \int^{\infty}_{0} p^2 dp \int^{\infty}_{0} dE\, \delta_\gamma (E, \Ep) e^{-E/T} }
\end{equation}
can be non-Maxwellian. We consider the case of a Maxwellian 
energy distribution, which is relevant for the deuteron 
distribution in metals that is discussed in the next section. A 
more general Fermi distribution, relevant for high-density 
environments, yields analogous effects and, in particular, a 
power-law tail for the momentum distribution. 

For the sake of concreteness let us consider a Coulombian 
collisional cross section, $ \sigma_{coll} \, =\, e^4/ \Ep^2 $; 
the resulting dispersion-relation width is
 \be
 \gamma =\hbar n \frac{e^4}{\Ep^2} \sqrt{\frac{2 E}{m}} =
 \left(\frac{\scaleE}{\Ep}\right)^2 \times \sqrt{\frac{E}{\scaleE}} \times \scaleE
                      \quad ,
 \ee
 where the collisional velocity $v_{coll} = \sqrt{2E/m}$ has been
 used, $n$ is the density and $m$ the mass of the colliding particles, deuterons
 in the present case. For
 convenience we have defined the energy scale
 \be
 \label{eq:energyScale}
  \scaleE = \left(\frac{m_e}{m}\right)^{1/5}
  \left(\frac{n}{n_0}\right)^{2/5} E_0 = \left(\frac{m_p}{m}\right)^{1/5}
    \left(\frac{n}{n_0}\right)^{2/5}  3.02649 \mathrm{\ eV} \quad,
\ee
 where $m_e$ and $m_p$ are the electron and proton masses, $E_0=(1/2)\alpha^2 m_e c^2$ is the Rydberg
 energy,  $n_0=(2 a_0)^{-3}= 0.843542\times 10^{24}$~cm$^{-3}$ a
 reference density with $a_0$ the Bohr radius.

Then the dispersion relation can be written as
\begin{equation}
   \delta_\gamma (E, \Ep)
  = \frac{1}{\scaleE\pi}   \frac{(\Ep/\scaleE)^2 \sqrt{E/\scaleE}}
   {(\Ep/\scaleE)^4(E/\scaleE - \Ep/\scaleE)^2 +  E/\scaleE}
   \quad ,
\label{eq:disprel}
\end{equation}
and the  momentum distribution is conveniently expressed in terms
of the scaled variable  $y=\Ep/(T \scaleE^5)^{1/6}$: \be
   \Phi\left(\frac{\Ep}{(T \scaleE^5)^{1/6}}, \frac{T}{\scaleE} \right) d\Ep  =
                                  N\left(\frac{T}{\scaleE}\right) \times y^{5/2} dy
                                   \int_0^\infty dx \frac{\sqrt{x}
e^{-x}}{x+y^4\left(y-x\left(\frac{T}{\scaleE}\right)^{5/6}\right)^2}
\quad ,
                                  \label{eq:AdoptMomentDisFun}
 \ee
 with the normalization
 \be
        N^{-1}\left(\frac{T}{\scaleE}\right)    =  \int^{\infty}_{0}dy y^{5/2} \int_0^\infty
        dx \frac{\sqrt{x} e^{-x}}{x+y^4\left( y - x \left( \frac{T}{\scaleE} \right)^{5/6}
\right)^2}
 \label{eq:NormAdoptMomentDisFun}
      \quad .
\ee

This distribution should be compared to the Maxwellian one
obtained when $\delta_{\gamma}(E,\Ep) = \delta(E-\Ep)$
\begin{equation}\label{eq:distrMax}
 \Phi_M\left(\Ep/T\right)d\Ep = dz  \frac{2}{\sqrt{\pi}} \sqrt{z}
 e^{-z}
\end{equation}
with $z=\Ep/T$.

Using the scaled variable  $\Ep / (T^{1/6}\scaleE^{5/6})$, the
distribution depends only on the adimensional parameter
$T/\scaleE$. Both the distribution $\Phi(\Ep)$\/ and the thermal
mean $\left< \sigma\, v_{rel} \right>$\/ can  be obtained
numerically. We have found an analytical approximation for the
important physical regime
 \be
     \label{eq:QEDistFunCond1}
     T \ll \scaleE
 \ee
that is sufficiently accurate and allows a better analysis. In
fact, since contributions to the integral for energies $x>1$ are
exponentially suppressed, Eq.~(\ref{eq:QEDistFunCond1}) implies
that the limit $T/\scaleE \to 0$ in
Eqs.~(\ref{eq:AdoptMomentDisFun}) and
(\ref{eq:NormAdoptMomentDisFun}) is well-defined. In this limit
the distribution becomes:
 \be
  \Phi_0\left(\frac{\Ep}{(T \scaleE^5)^{1/6}}\right) d\Ep
  \equiv
  \Phi\left(\frac{\Ep}{(T \scaleE^5)^{1/6}}, 0 \right) d\Ep  =
    N_0 y^{5/2} dy
    \int_0^\infty dx \frac{\sqrt{x} e^{-x}}{x+y^6} \quad ,
           \label{eq:AdoptMomentDisFunApprox}
  \ee
 where
   \be
   N_0^{-1} = \int^{\infty}_{0}dy y^{5/2}
              \int^{\infty}_{0} dx  \frac{ \sqrt{x} \; e^{-x}}{y^6 + x} =
              \frac{\pi\sqrt{2}\Gamma(13/12)}{3(1+\sqrt{3})} =0.51946 \quad .
   \ee
Under condition (\ref{eq:QEDistFunCond1}) the distribution 
depends on the temperature $T$ and density $n$ only through the 
single scale parameter $T^{1/6} \scaleE^{5/6}\sim T^{1/6}
 n^{1/3}$, which replaces the temperature $T$ of the
 Maxwell-Boltzmann distribution.

An estimate of corrections to this limiting behavior can be
obtained by considering the leading corrections to the
normalization integral
\begin{eqnarray}
  N^{-1}\left(T / \scaleE \right)
  &=& N_0^{-1} +
  \frac{ 5\pi\Gamma(23/12) }{ 9 ( \sqrt{2}+\sqrt{6} ) }  \left( T / \scaleE \right)^{5/6}
       + {\cal O} \left( (T / \scaleE)^{5/3}  \right) \\
  &=& 0.519457 +
      0.437082 \left( T / \scaleE \right)^{5/6}
      + {\cal O} \left( (T / \scaleE)^{5/3}  \right)
  \quad .
\end{eqnarray}

In the limit of small and large $\Ep$ the momentum 
distribution~(\ref{eq:AdoptMomentDisFunApprox}) behaves like
\begin{eqnarray}
  \lim_{\Ep\to 0 }  \Phi_0(y)dy &=& 
  N_0\sqrt{\pi} y^{5/2} dy \left( 1- \sqrt{\pi} y^3 + 2 y^6 + {\cal O}(y^{9}) \right)\\
  \lim_{\Ep\to\infty} \Phi_0(y)dy  &=& 
  N_0\frac{\sqrt{\pi}dy}{2 y^{7/2}}
  \left( 1- \frac{3}{2 y^6} +\frac{15}{4 y^{12}} + {\cal
  O}(y^{-18})\right) \quad .
\end{eqnarray}

Approximations that retain the leading and next-to-leading
asymptotic behaviors of the distribution are
\begin{eqnarray}
   \Phi_{l}(y)dy &=& N_0 \sqrt{\pi} y^{5/2} dy \frac{1}{1+2 y^6}\\
   \Phi_{nl}(y)dy  &=& N_0 \sqrt{\pi} y^{5/2} dy \frac{1+ y^6}{1+\sqrt{\pi} y^3+ 5y^6+ 2 y^{12}} \quad .
\end{eqnarray}
The linear combination of the two approximations that maintains
the normalization, $\int_0^\infty dy (c \Phi_{l}(y)+
(1-c)\Phi_{nl}(y))=\int_0^\infty dy \Phi_0(y)$, reproduces
$\Phi_0(y)$ within 2\%.

The most important lesson we learn from this analysis is that the
quantum broadening due to plasma effects  produces a distribution
whose typical scale is $(T\scaleE^5)^{1/6}$ instead of $T$: these
two scales can be very different. For instance, deuterons, $m =2 
m_p$, at the density $n=4.38\times 10^{23}$ cm$^{-3}$ have the 
energy scale $\scaleE \approx 2.03$~eV. Therefore, at 
temperatures $T_1=2.44\times 10^{-2}$~eV and $T_2=0.109$~eV, 
which fulfil condition (\ref{eq:QEDistFunCond1}), the scales of 
the modified distributions are $\scaleE^{5/6} T_1^{1/6} = 
0.770245$~eV and $\scaleE^{5/6} T_2^{1/6} =  0.988481$~eV, 
respectively.

This large shift of the particle distribution towards higher 
energies is demonstrated in Fig.~\ref{fig:QeDistFun}, where we 
show $\Phi_0(\Ep)$ from Eq.~(\ref{eq:AdoptMomentDisFunApprox}) 
compared to the Maxwell-Boltzmann distribution, 
Eq.~(\ref{eq:distrMax}), for the two temperatures $T_1$ and 
$T_2$; the target deuteron density of Ref.~\cite{Raiola:02} has 
been used as density $n$ of colliding particles: $n = 4.38 \cdot 
10^{23}$~cm$^{-3}$ .

\section{The low-energy {\em d(d,p)t} reaction rate. }

The {\em d(d,p)t} fusion reaction rate has been recently measured
using deuterated metal targets in the 4-20 keV energy range
\cite{Raiola:02, Raiola:02a, Bonomo:03, Raiola:04, Raiola:05}. At 
low energy these experiments have found a considerable higher 
reaction rate than the corresponding one measured using gas 
targets. Low-energy enhancements are usually explained in terms 
of electron screening; however, the electron screening potential 
$U_e$ that would reproduce these measurements in deuterated 
metals is of the order of hundreds of eV: this potential is much 
higher than the adiabatic estimate for the maximal screening 
potential $U_e \leq\;$ 28 eV.

The thermal motion of the target atoms is another mechanism
capable of increasing the reaction rate; however, the Maxwellian
momentum distribution at the experimental temperatures gives a
negligible effects~\cite{Fi:03, Starostin:2003next}.

The observation that large enhancements have been observed in
deuterated metals but not in insulators~\cite{Bonomo:03,Raiola:05}
has suggested a possible explanation based on effects of the 
plasma of electrons in the metal~\cite{Raiola:04, Raiola:05}. 
This simplified model with quasi-free valence electrons predicts 
an electron screening distance of the order of the Debye length 
$R_{Deb} = \sqrt{k_b T/(4\pi n_{\mathrm{eff}} (Ze)^2)}$, where 
$n_{\mathrm{eff}}$ is the effective density of valence electrons 
that can be treated as quasi-free. This approach reproduces both 
the correct size of the screening potential $U_e$ and its 
dependence on the temperature: $U_e \propto 
T^{-1/2}$~\cite{Bonomo:03, Raiola:05}.

The problem with this explanation is that the resulting 
$R_{Deb}$, for the actual experimental conditions, is about ten 
times smaller than the Bohr radius $a_0$; the mean number of 
quasi-free particles in the Debye sphere $N_{Deb}$, the so called 
Debye number~\cite{Ichimaru:1992}, is, therefore, much smaller 
than one: $N_{Deb} = n_{\mathrm{eff}} (4 \pi/3) 
R_{\mathrm{Deb}}^3  \approx (4 \pi/3) n_{\mathrm{eff}} (a_0/10)^3 
\approx 3 \cdot 10^{-5}$. The picture of the Debye screening, 
which should be a cooperative effect with many participating 
particles  ($R_{\mathrm{Deb}}$ should be at least greater than 
the Wigner Seitz radius, which is of the order of the Bohr 
radius), seems not to be applicable and the observed increase of  
the  {\em d(d,p)t} reaction rate still missing a consistent 
explanation. An additional technical inconsistency in the Debye 
screening explanation~\cite{Bonomo:03,Raiola:05} is the use of a 
non-degenerate formula for the screening radius in a situation 
where the electrons are degenerate.

In this context, we apply the analysis presented in the previous
section to the discussion of a recent interesting tentative
explanation of these puzzling experimental rates, which is based
on the quantum-tail
effect~\cite{Starostin:2003next,Starostin:2003next2}.

Since it is a good approximation of the experimental situation to 
consider the projectile distribution monoenergetic with a sharp 
relation between energy and momentum 
$\delta_{\gamma}(E_{beam},\Ep)=\delta(E_{beam}-\Ep)$, the 
relative velocity is function only of $\Ept $ and of the angle 
$\vartheta$ between $\mathbf p_t$ and the beam: $v_{rel}^2 = 2 
(E_{beam} + \Ept-2\,\sqrt{E_{beam}\Ept}\cos\vartheta )/m$. 
Following the analysis of the previous section, target particles 
have a Boltzmann-Gibbs energy distribution with a relation 
between energy and momentum that is broadened by plasma quantum 
effects and it is  given by Eq.~(\ref{eq:disprel}). The effective 
momentum distribution of the target particles is, therefore, not 
Maxwellian but given by Eq.~(\ref{eq:AdoptMomentDisFun}). 
Substituting this distribution of target particles, the sharp 
monoenergetic distribution for projectile particles, and the 
above relation for the relative velocity in 
Eq.~(\ref{eq:ChargPartReactRate}) the reaction rate per particle 
becomes
\begin{eqnarray}
    \left<  \sigma\, v_{rel} \right>  &  =  &
    \int d^3\mathbf p_t\, \Phi(\mathbf p_t)\, \sigma\, v_{rel}\, =\,
    \label{eq:NucReacRate} \\
                                      &  =  &
     2 \pi m_D^{3/2}\,
      \int_{-1}^{+1} d\cos\vartheta \int_0^{\infty} dE \int_0^{\infty} d\Ept
      \sqrt{2 \Ept}\, \delta_{\gamma}(E, \Ept)\, e^{-E/k_bT}\  \\
      &=&  \frac{1}{2}\int_{-1}^{1} 
    d\cos\vartheta \int_0^{\infty} d\Ept \Phi(\Ept/(T\scaleE^5)^{1/6}, T/\scaleE) 
    \sigma\, v_{rel} \quad , \nonumber \\
       \sigma\, v_{rel}\; , \nonumber \\
    \sigma\, v_{rel}                  &  =  &
    \frac{4\,S(E_{cm})}{m_D\,  v_{rel}}\,
    \exp\left(-\pi \sqrt{\frac{2 E_G}{\mu v_{rel}^2}}\right)\; ,
    \nonumber
\end{eqnarray}
where $E_{cm} = \mu v_{rel}^2 /2 $ and $\mu = m/2$. Since 
condition $T\ll\scaleE$ is verified, we can use the simplified 
form in Eq.~(\ref{eq:AdoptMomentDisFunApprox})
\begin{equation}\label{eq:sigmaVapprox}
    \langle \sigma v_{rel}\rangle = \frac{1}{2}\int_{-1}^{1} 
    d\cos\vartheta \int_0^{\infty} d\Ept \Phi_0(\Ept/(T\scaleE^5)^{1/6}) \sigma\, v_{rel} \quad .
\end{equation}

As we have analyzed in the previous section, the momentum
distribution resulting from the quantum broadening with the chosen
collisional cross section has two main features: a peak at
energies of the order of $(T\scaleE^5)^{1/6}$ instead of $T$ and a
power-law tail that decreases as $\Ep^{-7/2}$ instead of the
exponential cut-off. It is, therefore, physically interesting to
separate the contributions from the peak, the high-momentum tail 
and the low-momentum part of the distribution, so that we can 
understand which feature(s) of the modified distribution give(s) 
important corrections to the rate. To this purpose we define the 
peak of the target momentum distribution as $0.54 
(T\scaleE^5)^{1/6}=\epsilon_{l}<\Ep<\epsilon_{h}=1.15(T\scaleE^5)^{1/6}$: 
this region includes about 50\% of the particles and we call the 
corresponding contribution to the reaction rate per particle 
$\left<  \sigma\, v_{rel} \right>_C$; the contributions from the 
low- ($\Ep<\epsilon_{l}$: about 10\% of the particles) and 
high-momentum ($\epsilon_{h}<\Ep$: about 40\% of the particles) 
parts are indicated with the subscript $L$ and $H$.
\begin{equation}\label{eq:SeparateCont}
\left<  \sigma\, v_{rel} \right>   =    \left<  \sigma\, v_{rel}
\right>_L\, +\,
                                               \left<  \sigma\, v_{rel} \right>_C\, +\,
                                               \left<  \sigma\, v_{rel} \right>_H
\end{equation}

We have used $S(E_{CM})\simeq S_0 = 43 $~keV~b (the error is
$\leq$ 6\% for $E_{beam} \leq 10$~keV): then the $\vartheta$
integral can be done analytically in terms of the incomplete
Gamma-Euler function; the remaining integrations have been
performed using the Gauss-adaptive method.

The resulting rate is shown in Fig.~\ref{fig:QeExpConf} as the
thin solid line. In the same figure are shown for comparison the
experimental data~\cite{Raiola:02} and the other theoretical
curves: the thick solid line shows the rate for bare nuclei, while
the dot-dashed line shows the rate with electron screening in the
adiabatic limit, which should provide an upper limit to the
screening potential, $f_e = \exp\left( \pi
\sqrt{\frac{E_G}{E_{cm}}}\, \frac{U_e}{2\,E_{cm}} \right)$.

As it is apparent from Fig.~\ref{fig:QeExpConf}, the quantum-tail
effect is in fact capable to produce a strong enhancement of the
reaction rate, but this effect starts only below $E_{beam}\sim
2$~keV; on the contrary the experimental excess starts at energies
two or three times higher.

\begin{figure}
 \begin{center}

        \includegraphics[scale=0.6]{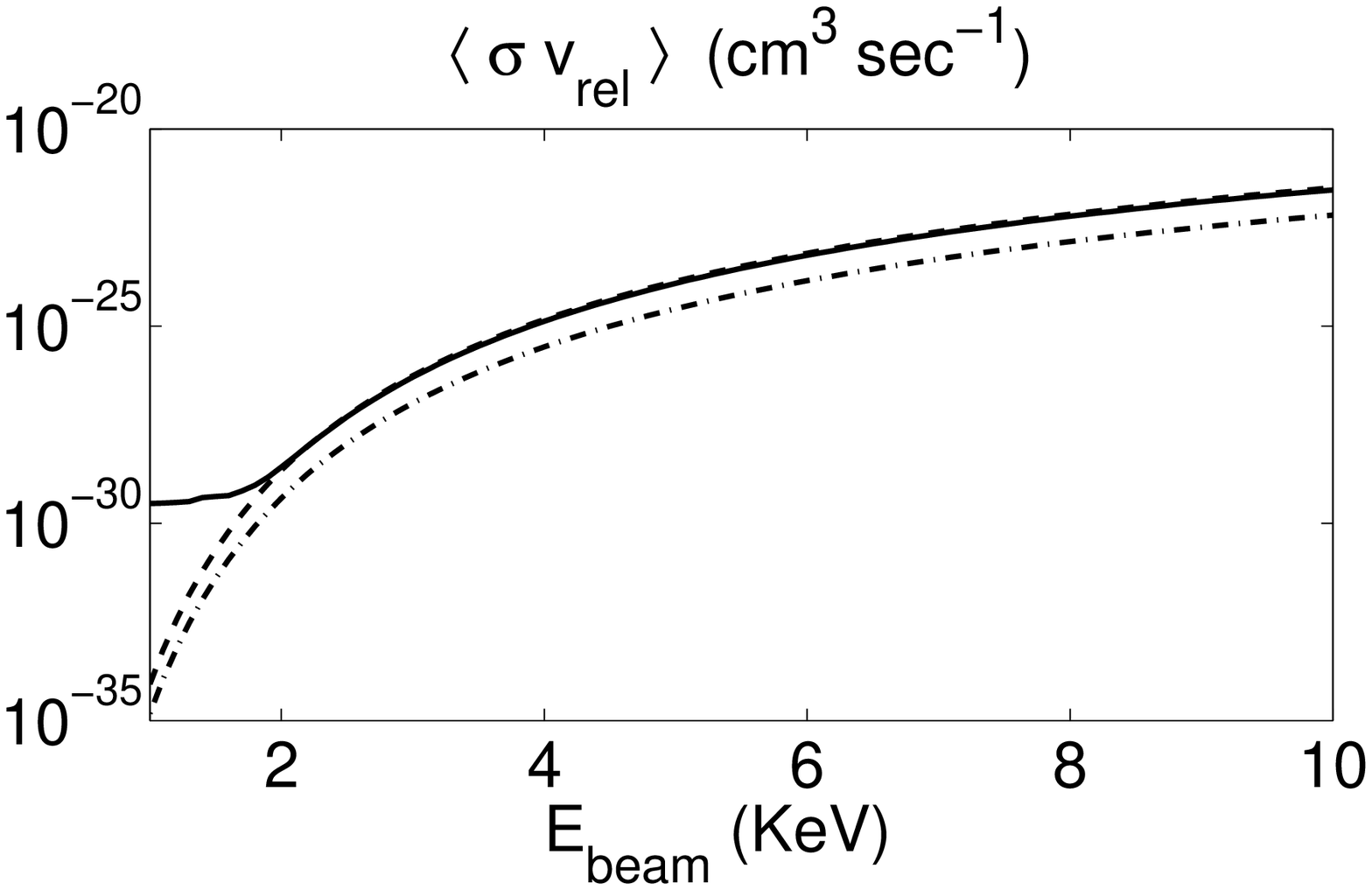}

 \end{center}
 \caption{\footnotesize Contributions to the reaction rate per particle
   $\left<  \sigma\, v_{rel} \right>$ coming from the three different
   regions of the target momentum distribution: low-momentum region (dash-dotted
   line), central or peak region (dashed line), and high-momentum
   or tail
   region (solid line).}
 \label{fig:QeSeparatCont}
 \end{figure}

From the analysis of the separate contributions of the three
regions (low, central or peak, and high) of the target momentum
distribution (see Fig.~\ref{fig:QeSeparatCont}), we observe that
the increase of $\left< \sigma\, v_{rel} \right>$ at low energies
($E_{beam}\sim 2$~keV), shown in Fig.~\ref{fig:QeExpConf}, is
caused mainly by the high momentum particles in the power-law
tail, the $\left< \sigma\, v_{rel} \right>_H$ term: the peak  and 
the low-momentum region seem not contribute to this increment in 
the present situation.

\section{Conclusion}
\begin{figure}
 \begin{center}
        \includegraphics[scale=0.8]{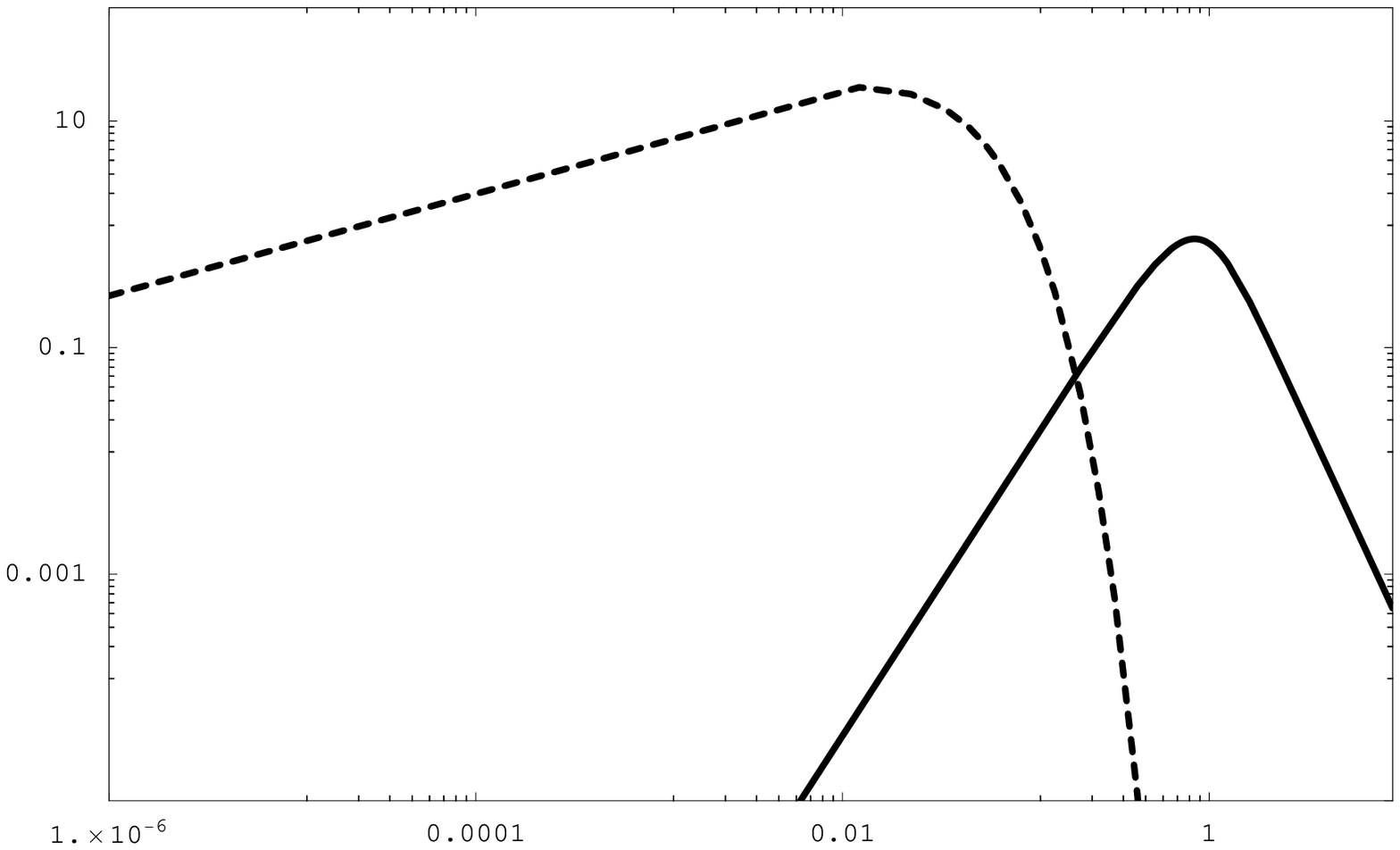}
        \hfill
        \includegraphics[scale=0.8]{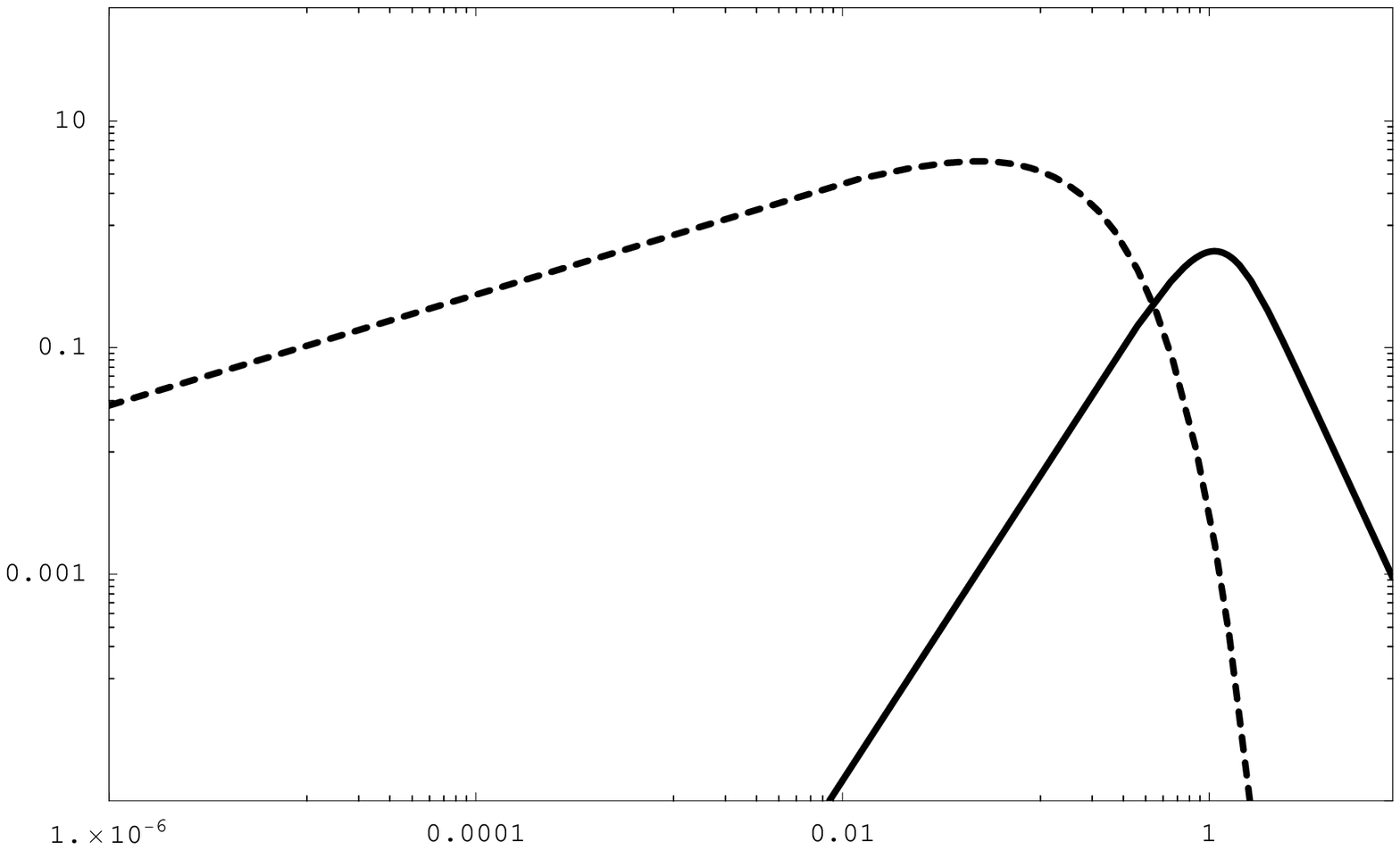}
 \end{center}
 \caption{The momentum distribution function $\Phi_0(\Ep)$ in
 eV$^{-1}$ (full line) and the Maxwellian one (dashed line) as functions of
 $\Ep$ for two different temperatures $T_1=10\;^0$C~$ = 0.0244$~eV
 (upper panel) and $T_2=1000\;^0$C~$= 0.109$~eV (lower panel) with density $n=4.38\times 10^{23}$ 
cm$^{-3}$.
         }
 \label{fig:QeDistFun}
 \end{figure}

We have studied the effects of the quantum broadening of the
relation between energy and momentum due to a specific collisional
cross section. An important effect is the considerable
modification of the resulting momentum distribution: \\
(1) the central part of the distribution is shifted from $T$ to
$(T \scaleE^5)^{1/6}$ where the energy scale $\scaleE$,
Eq.~(\ref{eq:energyScale}), grows as $n^{2/5}$ with the density
$n$; \\
(2) the high-momentum tail decreases as a power $\Ep^{-7/2}$
instead of having an exponential cut-off (see
Fig.~\ref{fig:QeDistFun}).

We have applied this quantum-tail effect to  nuclear fusion
processes between charged particles at sub-barrier energies of the
order of few keV and compared our results with the experimental
data relative to the {\em d(d,p)t} reaction with deuterated
target.

\begin{figure}
 \begin{center}

        \includegraphics[scale=0.9]{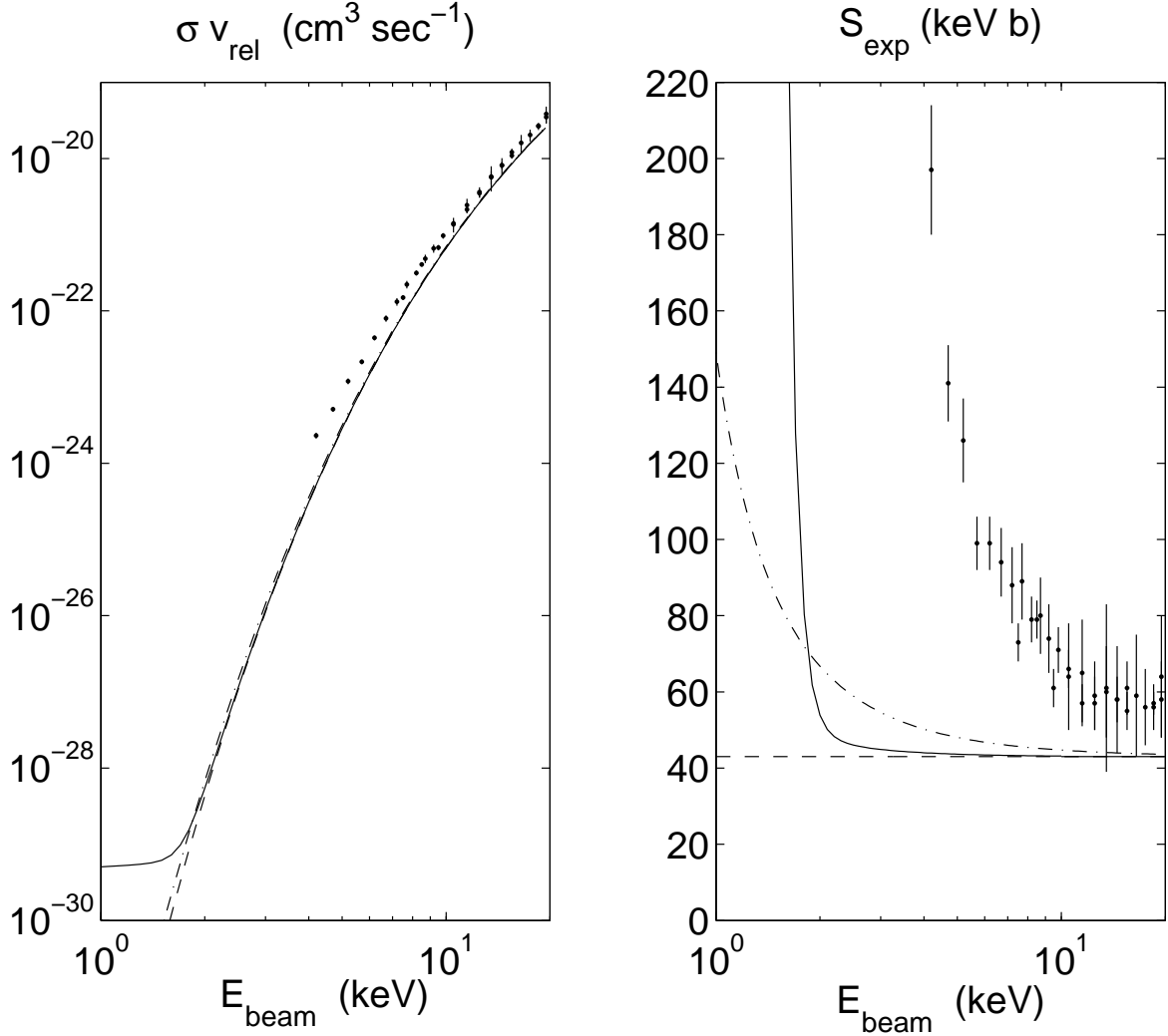}

 \end{center}
 \caption{Reaction rate per particle $\left(  \sigma\, v_{rel}
 \right)_{T=0}$ (left panel) and the corresponding astrophysical factor
          $S = \left(  \sigma\, v_{rel}  \right)_{T=0} \frac{m_D}{4} \sqrt{\frac{2 E_{beam}}{m_D}}
           \exp\left( \pi \sqrt{\frac{2\, E_G}{E_{beam}}} \right)$ (right panel) for the
           $d(d,p)t$ reaction  as function of the beam energy. The experimental
           data~\cite{Raiola:02} are compared with three theoretical
           curves: bare nuclei (dashed line),  screened
           nuclei with the adiabatic potential $U_e=28$~eV
          (dot-dashed line), and  our calculation that includes the
          quantum-tail thermal effect (solid line).
       }
 \label{fig:QeExpConf}
 \end{figure}

Our calculation shows that the quantum-tail effect produces an
important increment of the observed reaction rate enhancement at
very low energies ($\sim 2$~keV). However, this mechanism cannot
reproduce the experimental rate for deuterium, which has been
found to increase already at higher energies ($\sim 6-8$~keV), as
shown in Fig.~\ref{fig:QeExpConf}.

We have also analyzed more in details the effect of the modified
momentum distribution on the reaction rate by breaking up the 
contributions from target particle in three regions: the 
low-momentum, the central or peak, and the high-momentum region. 
The strong enhancement of the rate is due essentially to the 
particles in this last region: the high-momentum power-law tail 
of the distribution.

We are extending our results by using other collisional cross
sections and investigating the temperature dependence of the
mechanism.

\newpage

\begin{thebibliography}{Ga:67}
%
\bibitem{Ga:67}
V.~M.~Galitskii and V.~V.~Yakimets,
{\it JEPT} {\bf 24}, 637 (1967).
%
\bibitem{Coraddu:1998yb}
  M.~Coraddu, G.~Kaniadakis, A.~Lavagno, M.~Lissia, G.~Mezzorani and P.~Quarati,
  %``Thermal distributions in stellar plasmas, nuclear reactions and solar
  %neutrinos,''
  Braz.\ J.\ Phys.\  {\bf 29}, 153 (1999)
  [arXiv:nucl-th/9811081].
  %%CITATION = NUCL-TH 9811081;%%
 %
\bibitem{St:00}
A.~N.~Starostin, V.~I.~Savchenko, and N.~J.~Fisch, Phys. Lett. A  
{\bf 274}, 64 (2000).
%
\bibitem{St:02}
A.~N.~Starostin, A.~B.~Mironov, N.~L.~Aleksandrov, J.~N.~Fisch, 
and R.~M.~Kulsrud,  Physica A {\bf 305}, 287 (2002).
%
\bibitem{Lissia:2005em}
  M.~Lissia and P.~Quarati,
  %``Nuclear astrophysical plasmas: ion distribution functions and fusion
  %rates,''
  Europhys.\ Lett.\  {\bf 36}, 211 (2005)
  [arXiv:astro-ph/0511430].
  %%CITATION = ASTRO-PH 0511430;%%
%
\bibitem{Starostin:2003next}
%\bibitem{Coraddu:2004cp}
  M.~Coraddu, M.~Lissia, G.~Mezzorani, Y.~V.~Petrushevich, P.~Quarati and A.~N.~Starostin,
  %``Quantum-tail effect in low energy d+d reaction in deuterated metals,''
  Physica A {\bf 340}, 490 (2004)
  [arXiv:nucl-th/0401043].
  %%CITATION = NUCL-TH 0401043;%%
%
\bibitem{Starostin:2003next2}
%\bibitem{Coraddu:2004ku}
  M.~Coraddu, G.~Mezzorani, Y.~V.~Petrushevich, P.~Quarati and A.~N.~Starostin,
  %``Numerical modelling of the quantum-tail effect on fusion rates at low
  %energy,''
  Physica A {\bf 340}, 496 (2004)
  [arXiv:nucl-th/0402025].
  %%CITATION = NUCL-TH 0402025;%%
%
\bibitem{Raiola:02}
F.~Raiola {\it et al.}, Eur.\ Phys.\ J. {\bf A 13}, 377 (2002).
%
\bibitem{Raiola:02a}
F.~Raiola {\it et al.}, Phys.\ Lett.  {\bf B 547}, 193 (2002).
%
\bibitem{Bonomo:03}
C.~Bonomo {\it et al.},  Nuc.\ Phys. {\bf A 719}, 37c (2003).
%
\bibitem{Raiola:04}
F.~Raiola {\it et al.}, Eur.\ Phys.\ J. {\bf A 19}, 283 (2004).
%
\bibitem{Raiola:05}
F.~Raiola {\it et al.}, J.\ Phys.\ G {\bf 31}, 283 (2005).
%
\bibitem{Fi:03}
G.~Fiorentini, C. Rolfs, F. L. Villante and B. Ricci, Phys.\ Rev. 
C {\bf 67},  014603 (2003) [arXiv:astro-ph/0210537].
%
\bibitem{Ichimaru:1992}
S. Ichimaru, {\it Statistical plasma physics}, (Addison-Wesley, 
USA 1992).
%
\end{thebibliography}
\end{document}